\documentclass[12pt,aps,prd,preprint,superscriptaddress]{revtex4}

\usepackage{epsfig}
\usepackage{dsfont}
\usepackage{amssymb}
\usepackage{amsmath}
\usepackage{amsfonts}
\usepackage{graphics}

\newcommand{\mtrx}[2]{\left(\begin{array}{#1} #2 \end{array}\right)}
\newcommand{\I}[0]{{\rm i}}
\newcommand{\eps}{\varepsilon}
\newcommand{\Refs}{Refs.}
\newcommand{\Ref}{Ref.}
\newcommand{\Sec}{Sec.}
\newcommand{\tab}{Tab.}
\newcommand{\eq}{Eq.}

\newcommand{\fig}{Fig.}

\newcommand{\ie}{\emph{i.e.}}
\newcommand{\eg}{\emph{e.g.}}

\newcommand{\etc}{\emph{etc.}}
\newcommand{\CP}{\emph{CP}}
\newcommand{\etal}{\emph{et al.}}

\newcommand{\EE}[1]{\cdot 10^{#1}}

\newcommand{\capenv}[4]{%
	\begin{#1}
		#3
		\caption{\emph{#4}}
		\label{#2}
	\end{#1}%
}

\DeclareMathOperator{\diag}{diag}

\begin{document}

\title{Non-standard interactions using the OPERA experiment}

\author{Mattias Blennow}
\email[]{blennow@mppmu.mpg.de}
\affiliation{Max-Planck-Institut f\"ur Physik (Werner-Heisenberg-Institut),
F\"ohringer Ring 6, 80805 M\"unchen, Germany}

\author{Davide Meloni}
\email[]{meloni@fis.uniroma3.it}
\affiliation{Dipartimento di Fisica, Universit\'a di Roma Tre and INFN Sez.~di Roma Tre,
via della Vasca Navale 84, 00146 Roma, Italy}

\author{Tommy Ohlsson}
\email[]{tommy@theophys.kth.se}
\affiliation{Department of Theoretical Physics, School of Engineering Sciences, Royal Institute of Technology -- AlbaNova University Center,
Roslagstullsbacken 21, 106 91 Stockholm, Sweden}

\author{Francesco Terranova}
\email[]{Francesco.Terranova@cern.ch}
\affiliation{Laboratori Nazionali di Frascati dell'INFN, Via E.Fermi 40, 00044 Frascati, Italy}

\author{Mattias Westerberg}
\email[]{mwesterb@kth.se}
\affiliation{Department of Theoretical Physics, School of Engineering Sciences, 
Royal Institute of Technology -- AlbaNova University Center,
Roslagstullsbacken 21, 106 91 Stockholm, Sweden}

\begin{abstract}
We investigate the implications of non-standard interactions on
neutrino oscillations in the OPERA experiment. In particular, we study
the non-standard interaction parameter $\eps_{\mu\tau}$. We show that
the OPERA experiment has a unique opportunity to reduce the allowed
region for this parameter compared with other experiments such as the
MINOS experiment, mostly due to the higher neutrino energies in the CNGS beam compared to the NuMI beam.
We find that OPERA is mainly sensitive to a
combination of standard and non-standard parameters and that a
resulting anti-resonance effect could suppress the expected number of
events. Furthermore, we show that running OPERA for five years each
with neutrinos and anti-neutrinos would help in resolving the
degeneracy between the standard parameters and
$\eps_{\mu\tau}$.  This scenario is
significantly better than the scenario with
a simple doubling of the statistics by running with neutrinos for ten
years.

\end{abstract}


\preprint{MPP-2008-35}
\preprint{RM3-TH/08-7}
\maketitle

\section{Introduction}
\label{sec:intro}

Neutrino oscillation physics has definitively entered the era of
precision measurements of the fundamental neutrino parameters such as
the neutrino mass squared differences (\ie, $\Delta m_{31}^2$ and
$\Delta m_{21}^2$) and the leptonic mixing parameters (\ie,
$\theta_{12}$, $\theta_{13}$, $\theta_{23}$, and $\delta$). In
particular, the Super-Kamiokande, SNO, KamLAND, K2K, and MINOS
experiments have given valuable information on these parameters
\cite{Hosaka:2006zd,Ahmed:2003kj,Oblath:2007zz,Ahn:2006zza,Michael:2006rx,:2008ee}.

The precision measurements open up the possibility to investigate if
neutrino flavor transitions are governed by neutrino oscillations only
or if they are, in the next-to-leading order, a combination of
neutrino oscillations and some other new physics mechanism. However,
to leading order, there exist clear evidences that neutrino
oscillations constitute the underlying physical model for neutrino
flavor transitions.  The next-to-leading order mechanism could \eg~be
non-standard interactions (NSIs), mass varying neutrinos, neutrino
decay, neutrino decoherence, \etc~or some combination thereof.

In this work, we will study NSI effects at
the OPERA experiment \cite{Guler:2000bd}, which is an experiment that
consists of a massive lead/emulsion target (the OPERA detector)
located at LNGS in Gran Sasso, Italy, receiving its
neutrino beam, originally consisting almost exclusively of $\nu_\mu$,
from CERN in Geneva, Switzerland. The baseline length is approximately 732~km and the CNGS
$\nu_\mu$ beam has an average neutrino energy of $E_\nu \simeq 17$~GeV. 
The OPERA experiment is especially designed to observe
$\nu_\tau$ events from the $\nu_\mu \to \nu_\tau$ neutrino oscillation
channel. In fact, no previous experiment
has investigated this channel or observed neutrinos of a different flavor than that originally produced at the source (although the neutral-current measurements at SNO imply that solar $\nu_e$ have oscillated into a different flavor). Thus, the OPERA experiment presents a
unique opportunity to study direct appearance of
$\nu_\tau$~\cite{Acquafredda:2006ki}.
In this work, we will not try to describe the origin of the NSIs, but adopt a purely phenomenological point of view. In
particular, NSIs can modify the production, the propagation in
matter as well as the detection of the neutrinos. We will
concentrate on the simplified scenario in which NSIs only affect the
neutrino propagation.

Previously, investigations of NSIs that are of importance for this
work have been presented in the following papers: In
\Ref~\cite{Fornengo:2001pm}, a two-flavor neutrino analysis of the
so-called atmospheric neutrino anomaly has been performed, which
effectively bounds the NSI parameters in the $\mu$-$\tau$ sector,
$\eps_{\mu\tau} \simeq \eps$ and $\eps_{\tau\tau} \simeq \eps'$, to
$-0.03 \leq \eps \leq 0.02$ and $|\eps'| \leq 0.05$ at 99.73~\% confidence level.
Although these bounds may seem quite restrictive, it has been shown that at least the bound on $\eps_{\tau\tau}$ is severely weakened when considering the full three-flavor framework (allowing $\eps_{\tau\tau}$ to be of $\mathcal O(1)$ or larger, depending on the values of $\eps_{ee}$ and $\eps_{e\tau}$~\cite{Friedland:2005vy}). As will be shown later in this work, the limit that could be put by the OPERA experiment would be insensitive to whether the two- or three-flavor scenario is studied simply because of the relatively short baseline.
In addition, in \Ref~\cite{Ota:2002na}, the authors have come to the
conclusion that it would be possible to observe NSI effects at the
OPERA experiment (and the ICARUS experiment) if $\eps_{\mu\tau} \geq
{\cal O}(10^{-2})$.  Next, in \Ref~\cite{Ribeiro:2007jq}, the
Kamioka-Korea two detector setup has been investigated, which could also
give restrictions on the NSI parameters $\eps_{\mu\tau}$ and
$\eps_{\tau\tau}$.  Recently, in \Ref~\cite{EstebanPretel:2008qi}, a
study of the OPERA experiment (in combination with the MINOS experiment)
has been presented with the conclusion that it is not very sensitive to the NSI
parameters $\eps_{e\tau}$ and $\eps_{\tau\tau}$. However, it was found
that the $\nu_\tau$ sample is too small to be statistically
significant to improve the limits on the NSI parameter
$\eps_{\tau\tau}$.  Nevertheless, this analysis did not include a
study of the relevant $\eps_{\mu\tau}$ which, due to the energies and
the baseline involved in the OPERA experiment, is the only NSI parameter appearing to leading order in $L$ in the $\nu_\mu \to \nu_\tau$ flavor transition.

In general, neutrino oscillations and NSIs in terrestrial neutrino
experiments have been studied extensively in the literature, using the
neutrino factory project
\cite{NSInfstart,Ota:2001pw,GonzalezGarcia:2001mp,Gago:2001xg,
Huber:2001zw,Kopp:2007mi,Ribeiro:2007ud,NSInfstop} and other different
neutrino facilities (like super-beams and $\beta$-beams)
\cite{NSIotherstart,Adhikari:2006uj,Blennow:2007pu,Kopp:2007ne,NSIotherstop}
to assess the impact of the NSI effects in neutrino physics.

This work is organized as follows. In \Sec~\ref{analytic}, we will
present analytic considerations for the NSIs that we assume for the
OPERA experiment. In addition, we will comment on a sort of \emph{anti-resonance} effect that is in the vicinity of being detectable in
the OPERA experiment. Next, in \Sec~\ref{numset}, we will give the
numerical setup with the GLoBES
software~\cite{Huber:2004ka,Huber:2007ji} that we use for our
simulations of the OPERA experiment. Then, in \Sec~\ref{numres}, we
will show our numerical results for the OPERA experiment using
GLoBES. Finally, in \Sec~\ref{concl}, we will present a summary of the
work as well as our conclusions.

\section{Analytic considerations}
\label{analytic}

We consider effective non-standard interactions of the form
\begin{equation}
	\mathcal L_{\rm NSI} = - \frac{G_F}{\sqrt{2}}
	\sum_{\stackrel{f=u,d,e}{a=\pm 1}}
	  \eps^{fa}_{\alpha\beta}[\overline{f} \gamma^\mu (1+a\gamma^5) f]
	  [\overline{\nu_\alpha}\gamma_\mu(1-\gamma^5)\nu_\beta],
\end{equation}
where $f$ is summed over the matter constituents and the parameters
$\eps_{\alpha\beta}^{fa}$, which are the entries of a Hermitian matrix $\eps^{fa}$, give the strength of the NSIs.
In a manner completely analogous to the derivation of the normal matter effect,
these interactions will result in an effective addition
\begin{equation}
\label{eq:hamnsi}
	H_{\rm NSI} = V \mtrx{ccc}{
		\eps_{ee} & \eps_{e\mu} & \eps_{e\tau} \\
		\eps_{e\mu}^* & \eps_{\mu\mu} & \eps_{\mu\tau} \\
		\eps_{e\tau}^* & \eps_{\mu\tau}^* & \eps_{\tau\tau}}
\end{equation}
to the neutrino oscillation Hamiltonian in flavor basis, where
$\eps_{\alpha\beta} = \sum_{f,a} \eps_{\alpha\beta}^{fa} N_f/N_e$ and
$V = \sqrt{2}G_F N_e$. Notice that, apart from the bounds on
$\eps_{\mu\tau}$ and $\eps_{\tau\tau}$ given in the \Sec~\ref{sec:intro}, we
are not aware of any paper discussing direct bounds on the effective
parameters $\eps_{\alpha\beta}$. However, experimental limits on the
parameters $\eps_{\alpha\beta}^{fa}$ can be found in
\Refs~\cite{Davidson:2003ha,Abdallah:2003np}, which imply that
$|\eps_{e\mu}^{fa}| \leq {\cal O} (10^{-4})$ and $|\eps_{ee}^{fa}|
\leq {\cal O} (1)$ \cite{Ribeiro:2007ud}. Thus, we can assume that
the effective parameters $\eps_{ee}$ and $\eps_{e\mu}$ are bounded at
the same order of magnitude as their corresponding parameters
$|\eps_{\alpha\beta}^{fa}|$.

The full three-flavor Hamiltonian describing neutrino propagation in
matter is given by
\begin{equation}
\label{eq:fullham}
	H = \frac{1}{2E} U \diag(0,\Delta m_{21}^2,\Delta m_{31}^2) U^\dagger
		+ H_{\rm MSW} + H_{\rm NSI},
\end{equation}
where $U$ is the leptonic mixing matrix, $\Delta m_{ij}^2 =
m_i^2-m_j^2$, and $H_{\rm MSW}$ is the addition from the standard
matter effect. Due to the quite large neutrino energy $E_\nu = {\cal
O}(10)$~GeV and the relatively short baseline $L\simeq 732$~km, both
$\Delta m^2_{31} L / (2 E_\nu) \ll 1$ and $V L \ll 1$, where $V$ is
the matter potential $V \simeq 1.1 \EE{-13}$~eV in the Earth's
crust ($\rho \simeq 2.7 \, {\rm g/cm}^3$) \cite{Akhmedov:2004ny}.
Thus, neutrino oscillations will not have time to fully develop. As a
consequence, the main characteristics of the flavor transition
probabilities will be given by truncating the flavor evolution matrix
$S = \exp(-\I HL)$ at order $L$, resulting in
\begin{equation}
	S \simeq \mathds 1 - \I H L.
\end{equation}
The \emph{off-diagonal} neutrino transition probabilities are then given by
\begin{equation}
	P_{\alpha \beta} = |S_{\beta\alpha}|^2 \simeq |H_{\beta\alpha}L|^2.
\end{equation}
The \emph{diagonal} neutrino survival probabilities in this expansion
are given by the unitarity condition $P_{\alpha\alpha} = 1 -
\sum_{\beta\neq\alpha} P_{\alpha\beta}$. As can be observed from this
consideration, the transition probabilities will only be affected by
the corresponding NSI element (\ie, $P_{\alpha\beta}$ just depends on
the NSI element $\eps_{\beta\alpha}$), while the survival
probabilities depend on the two off-diagonal NSI elements associated
with the flavor (\eg, $P_{\mu\mu}$ is affected by $\eps_{e\mu}$ and
$\eps_{\mu\tau}$). As expected, the diagonal NSI parameters do not
enter at short baselines.  Clearly, this is not true in general and at
higher orders in $L$, where the NSI parameters will enter all of the
neutrino oscillation probabilities. As an example, the NSI parameter
$\eps_{e\tau}$ will enter the flavor evolution matrix $S_{\mu\tau}$ at
${\cal O}(L^2)$ and then to $\cal{O}$$(L^3)$ in the transition
probability $P_{\mu\tau}$ (unless there is no interference between the $L$ and $L^2$ terms).
From the above consideration, we can conclude that the NSI parameter
of most interest for the OPERA experiment is $\eps_{\mu\tau}$.
That the parameters $\eps_{e\tau}$ and $\eps_{\tau\tau}$ are not
important has been already shown in \Ref~\cite{EstebanPretel:2008qi}.

The main physics goal of the OPERA experiment is to actually observe
oscillations of $\nu_\mu$ into $\nu_\tau$. With the effects of
$\eps_{\mu\tau}$ included, the transition probability $P_{\mu\tau}$
is given by
\begin{equation}
\label{eq:pmutau}
P_{\mu\tau} = |S_{\tau\mu}|^2 = \left|c_{13}^2\,\sin(2\theta_{23})\frac{\Delta
m_{31}^2}{4E_\nu}+\eps_{\mu\tau}^* V\right|^2 L^2 + {\cal O}(L^3),
\end{equation}
where we have neglected the small mass
squared difference $\Delta m_{21}^2$. From this consideration follows that there is
a degeneracy between the standard neutrino oscillation parameters and
the NSI parameter $\eps_{\mu\tau}$ as scenarios with the same value of
$|c_{13}^2\,\sin(2\theta_{23})\Delta
m_{31}^2/(4E_\nu)+\eps_{\mu\tau}^* V|$ will lead to the same neutrino
oscillation probability. Even if the degeneracy is broken by the
energy dependence of the first term, we still expect some parameter
correlations when analyzing the outcome of an experiment. It is also
interesting to note that the $\mathcal O(L^2)$ contribution to $P_{\mu\tau}$
vanishes when
\begin{equation}
	\eps_{\mu\tau}^* = -c_{13}^2\, \frac{\Delta m_{31}^2}{4E_\nu\,V} 
	\sin(2\theta_{23})
\end{equation}
simply due to the fact that $S_{\tau\mu} = 0$ in this case. The
condition clearly shows that this can happen only for real
$\eps_{\mu\tau}$. We will use the term \emph{anti-resonance} to refer
to this scenario as it, in some sense, is the opposite of the
MSW-resonance: in the standard picture of neutrino oscillations, the
matter effects cancel the difference between the diagonal terms and
the effective mixing angle is maximal, whereas in the
situation with NSIs, the matter effects cancel the off-diagonal terms and the effective
mixing angle is minimal (\ie, zero). In a pure two-flavor scenario,
the anti-resonance is valid to all orders, while transitions can be
induced to higher order in $L$ by other off-diagonal elements in the
case of three-flavor oscillations. For the peak energy of $E_\nu \simeq
17$~GeV in the CNGS beam, the anti-resonance would occur for
$\eps_{\mu\tau} \simeq -0.3$ with the result that no $\nu_\tau$ events
would be observed. Note that a similar conclusion applies in the
case of inverted mass hierarchy, from which $\eps_{\mu\tau} \simeq
+0.3$ if $\Delta m_{31}^2 \to -\Delta m_{31}^2$ (neglecting the small
effect of $\Delta m^2_{21}$). This also applies to the case of anti-neutrinos, where we have $V \to -V$ and $\eps_{\alpha\beta} \to \eps_{\alpha\beta}^*$. In both cases, this also gives an estimate of the order of magnitude of the NSIs that OPERA will be sensitive to, as the expected number of $\nu_\tau$ events is low.

Finally, we want to mention that a similar effect could exist in the $\nu_\mu
\to \nu_e$ transition. In fact,
\begin{equation}
\label{eq:pmue}
P_{\mu e} = |S_{e \mu}|^2 = \left|\left[s_{23} \sin(2\theta_{13})
{\rm e}^{\I \delta} + \alpha c_{23} c_{13}
\sin(2\theta_{12})\right]\frac{\Delta m_{31}^2}{4E_\nu}+\eps_{e \mu}^*
V\right|^2 L^2 + \mathcal O(L^3),
\end{equation}
where $\delta$ is the standard \CP-violating phase in the unitary
leptonic mixing matrix, $\alpha = \Delta m_{21}^2/\Delta m_{31}^2$ is
the ratio between the mass squared differences, and we have neglected
a term proportional to $s_{13}\alpha$.
In this case, the external bounds on $\eps_{e\mu}$ are so stringent
that the term proportional to $\alpha$ is known to be larger. Thus,
an anti-resonance in this channel could only be due to an interplay
between the two standard terms if $\delta = \pi$.

\section{Numerical setup}
\label{numset}

The numerical simulations of the OPERA experiment were performed using
the GLoBES software~\cite{Huber:2004ka,Huber:2007ji}, which was
extended in order to accommodate the inclusion of NSIs through the
Hamiltonian presented in \eq~(\ref{eq:hamnsi}) with $\eps_{ee} =
\eps_{e\mu} = \eps_{e\tau} = 0$. The neutrino propagation in matter was
then described using the full three-flavor Hamiltonian in
\eq~(\ref{eq:fullham}).
In addition, the Abstract Experiment Definition Language (AEDL) file,
used to describe the OPERA experiment, was based on the results
presented in \Refs~\cite{Guler:2000bd,Komatsu:2002sz,Huber:2004ug}.
Unless stated otherwise, we have assumed a running time of five years with
$4.5\EE{19}$ protons on target per year, in accordance with the
OPERA experimental setup, and an effective mass of $1.65$~kton
\cite{Guler:2000bd}. Furthermore, the neutral- and charged-current
cross-sections were taken from
\Refs~\cite{Guler:2000bd,Messier:1999kj,Paschos:2001np}.
The CNGS neutrino spectra are substantially different from zero in the
interval between 1~GeV and 30~GeV (with a peak around $E_\nu \simeq 17$
GeV). Thus, we divided the signals and the corresponding backgrounds into 29
equally spaced energy bins, having checked that the numerical results
are stable if the number of energy bins is above the order of 10. For
the baseline length of the CNGS setup (approximately 732~km), the
matter density profile was assumed to be constant and equal to the
value at the Earth's crust, \ie, $\rho = 2.72~{\rm g/cm^3}$ (or $V
=1/1900~{\rm km}^{-1}$)~\cite{Dziewonski:1981xy}. In all simulations,
we have used a full three-flavor neutrino framework with central
values and 1$\sigma$ errors of the standard neutrino oscillation
parameters as given in \tab~\ref{tab:oscparams}. Normal mass
hierarchy, \ie, $\Delta m^2_{31}>0$, has been assumed if not stated otherwise.

Regarding the NSI parameters, we performed numerical simulations with
different simulated values, also taking into account the effects of possible
\CP-violating phases of the non-diagonal entries of the Hamiltonian in
\eq~(\ref{eq:hamnsi}). The priors set on the NSI parameters are
chosen according to \Ref~\cite{Davidson:2003ha}, except from
$\eps_{\tau\tau}$, which has further been constrained using atmospheric
neutrino data~\cite{Friedland:2005vy}.
\capenv{table}{tab:oscparams}{%
\centering
\begin{tabular}{|l|l|}
\hline
$\theta_{12} = 34.4^\circ \pm 1.7^\circ$ &
$\Delta m_{21}^2 = (7.59\pm 0.21) \EE{-5} \, {\rm eV}^2$
\\
\hline
$\theta_{13} = 4.8^\circ \pm 2.9^\circ$ &
$\Delta m_{31}^2 = (2.4 \pm 0.15) \EE{-3} \, {\rm eV}^2$ \\
\hline
$\theta_{23} = 45^\circ \pm 3.8^\circ$ & $\delta = \pi/2$ \\
\hline \end{tabular}%
}{%
The simulated values of the
standard neutrino oscillation parameters and the corresponding $1\sigma$ priors used in the simulations.
The central values of the parameters $\theta_{12}$ and $\Delta m_{21}^2$ were
inspired by the results of the KamLAND experiment \cite{:2008ee},
whereas the central values of the other parameters were inspired by
\Ref~\cite{Maltoni:2004ei}. We fixed the value of the
\CP-violating phase to $\pi/2$, with no consequences on our results for the 
$\nu_\mu\rightarrow \nu_\tau$ channel.
}
As a comparison, we also included the MINOS experiment, able to probe the $\nu_\mu \to \nu_e$ transition channel, in our simulations. As already mentioned in \Ref~\cite{EstebanPretel:2008qi}, different $L/E_\nu$ could in general be very useful in order to further
constrain some of the parameters of \eq~(\ref{eq:hamnsi}), since the relative importance of the standard and non-standard parts of the Hamiltonian is energy dependent. Our
numerical setup of the MINOS experiment follows that used in
\Ref~\cite{Blennow:2007pu}.

\section{Numerical results}
\label{numres}

In this section, we present the numerical results on the physics reach
of the OPERA experiment in constraining the new physics parameters
$\eps_{\alpha\beta}$. In all figures, we have combined both the $\nu_\mu \to
\nu_e$ and $\nu_\mu \to\nu_\tau$ channels for the OPERA experiment.

The results have been obtained by marginalizing over the parameters
$\Delta m^2_{31}$ and $\theta_{23}$ (if not stated otherwise), while keeping the parameters
$\Delta m_{21}^2$ and $\theta_{12}$ fixed, since they are irrelevant for
the $\nu_\mu \to\nu_\tau$ transition in the OPERA experiment. In
addition, the parameter $\theta_{13}$ was fixed, since it does not
affect the results. We also observed that
$\eps_{ee},\eps_{e\mu},\eps_{\mu\mu},$ and $\eps_{e\tau}$ do not
affect the results, which means that they are fixed to zero in the
rest of the work.

First, in \fig~\ref{fig:combination}, we present the sensitivity reach
for $\eps_{\mu\tau}$ with the OPERA experiment in combination with the
MINOS experiment (for a discussion on the sensitivity reach for
$\eps_{e\tau}$ and $\eps_{\tau\tau}$ for the same combination, see
\Ref~\cite{EstebanPretel:2008qi}).
\capenv{figure}{fig:combination}{%
	\begin{center}
		\includegraphics[width=0.75\textwidth,clip=true]{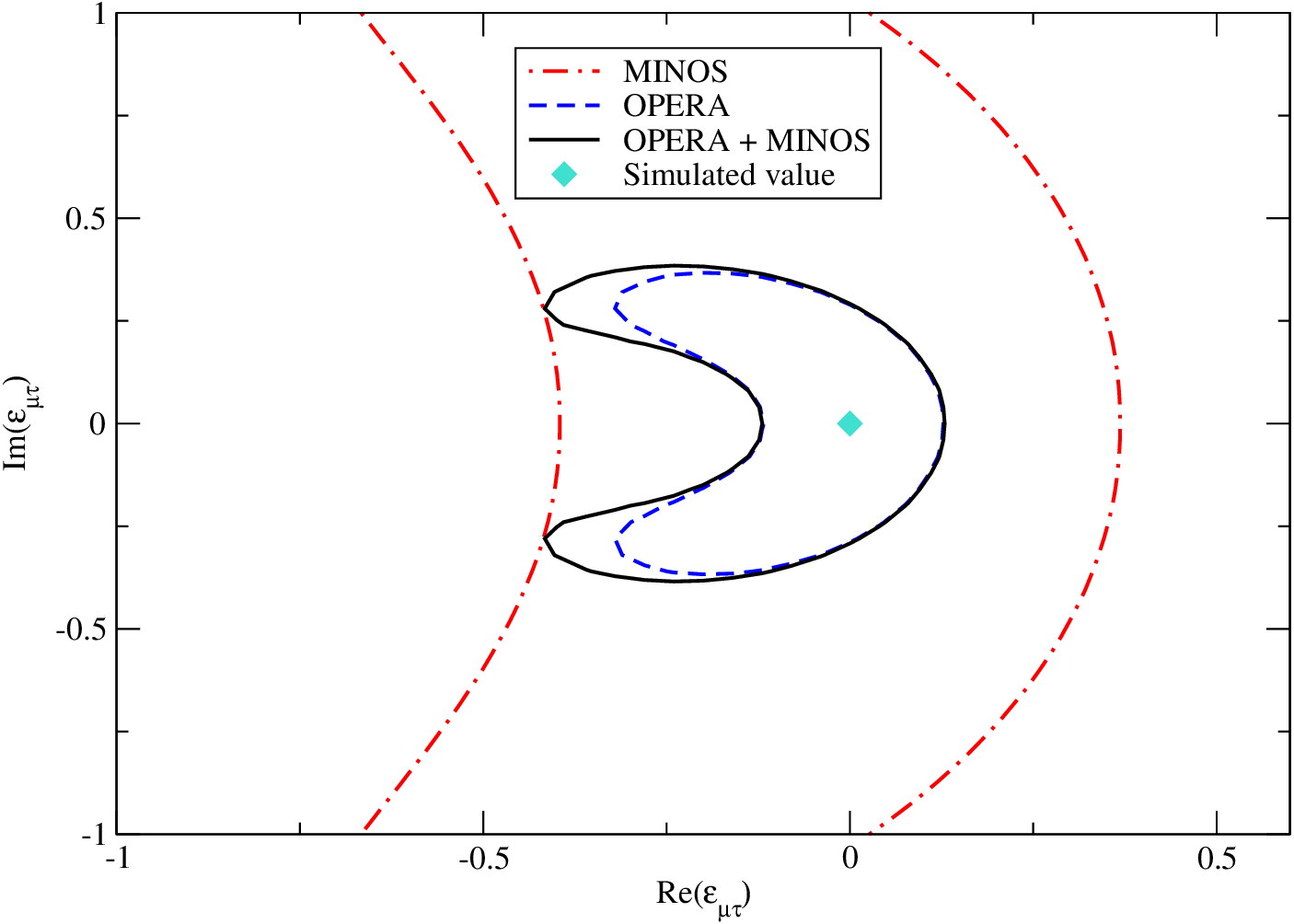}
	\end{center}%
}{%
Sensitivity for $\eps_{\mu\tau}$ at 95~\% confidence level (2 d.o.f.)
of the OPERA and MINOS experiments in the case of no NSIs (the input
values of the various $\eps_{\alpha\beta}=0$).}%
As can be observed in this figure, OPERA is far more sensitive to
$\eps_{\mu\tau}$ due to the higher neutrino energy than that in MINOS, which can therefore only
marginally improve the sensitivity. Thus, in the following we will
only consider the bounds which can be placed from OPERA itself.

Note that, as expected, a similar situation is also valid when
considering the detection of the $\nu_\mu \to \nu_e$ transition in
OPERA. In fact, as already stressed in \Sec~\ref{analytic},
$\eps_{\mu\tau}$ appears to leading order in $P_{\mu\tau}$
[\eq~(\ref{eq:pmutau})], but it is subleading in $P_{\mu e}$
[\eq~(\ref{eq:pmue})]. We verified that the inclusion of the $\nu_\mu
\to \nu_e$ channel does not affect the results on $\eps_{\mu\tau}$.

\subsection{Marginalization of $\boldsymbol{\eps_{\tau\tau}}$}

Figure \ref{fig:tautau-effect} shows the OPERA sensitivity in the
$|\eps_{\mu\tau}|$-$\eps_{\tau\tau}$ plane (left panel) as well as the
impact on the $\eps_{\mu\tau}$ sensitivity given different priors on
$\eps_{\tau\tau}$ (right panel).
\capenv{figure}{fig:tautau-effect}{%
	\begin{center}
		\includegraphics[width=0.49\textwidth,clip=true]{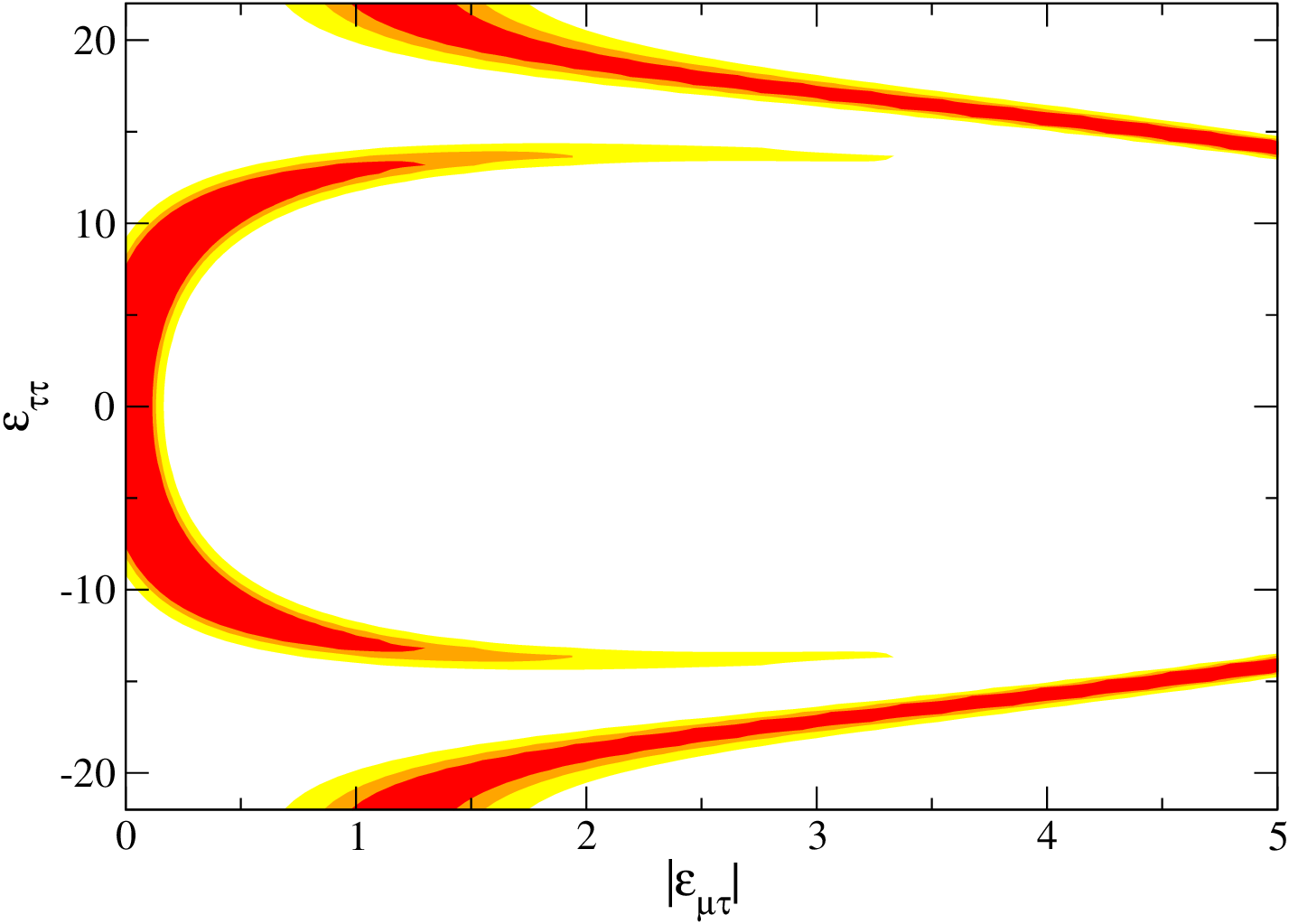}
		\includegraphics[width=0.49\textwidth,clip=true]{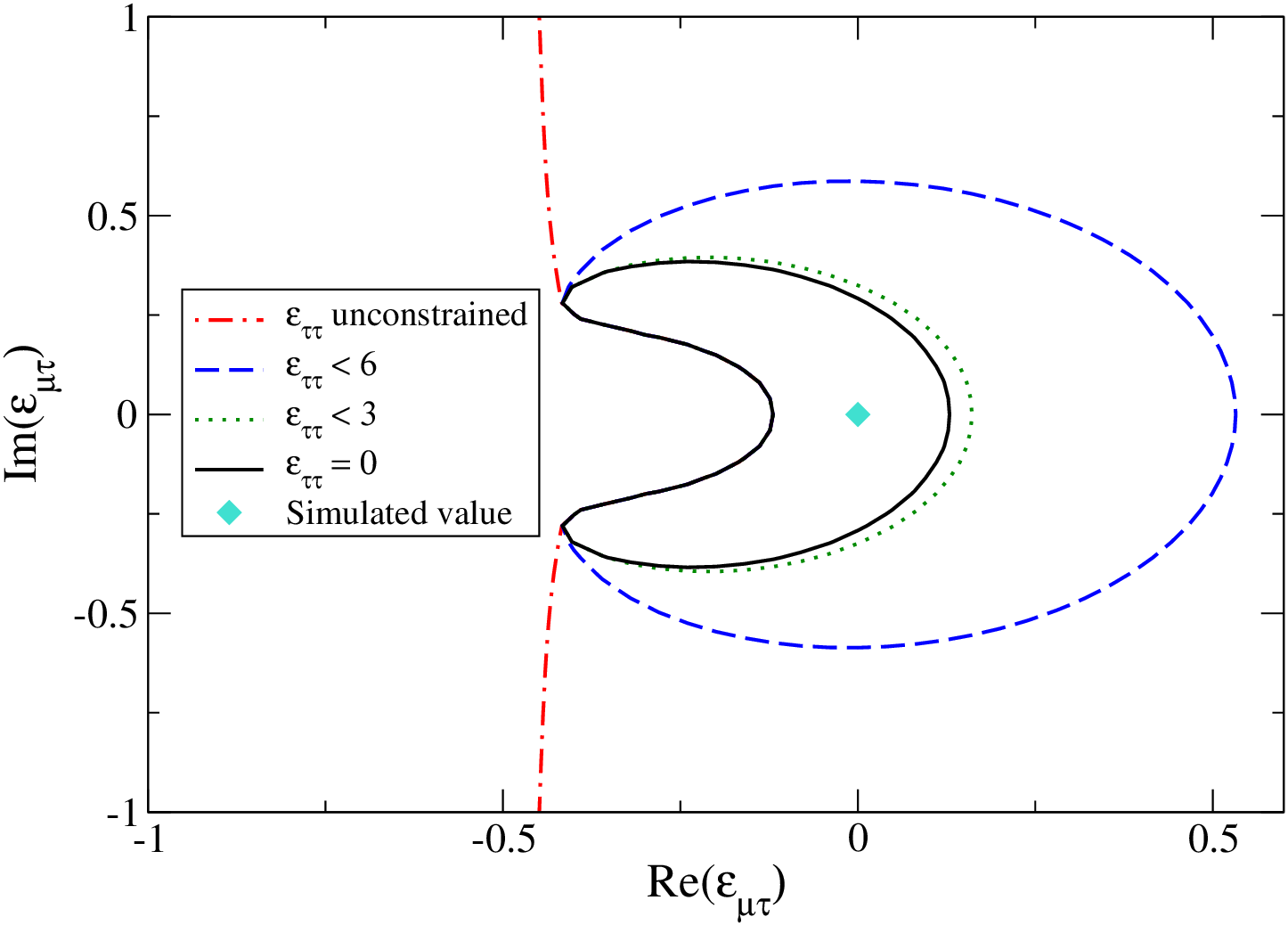}
	\end{center}
}{%
The left panel shows the NSI sensitivity of the OPERA experiment alone
for marginalized standard
neutrino oscillation parameters. The confidence levels (2~d.o.f.) are 90~\%, 95~\%, and 99~\%, respectively. The sensitivity contours for fixed standard neutrino oscillation parameters only differ slightly from this result. The right panel shows
how the sensitivity to $\eps_{\mu\tau}$ changes depending on the
$\eps_{\tau\tau}$ prior. The sensitivity contours in the right panel
are at 95~\% confidence level (2~d.o.f.) and the $\eps_{\tau\tau}$
priors are at $1\sigma$ level.}
As expected, the impact of $\eps_{\tau\tau}$ is small as long as any
reasonable prior is put. This is naturally related to the fact that
$\eps_{\tau\tau}$ enters only in higher order in the oscillation
probability $P_{\mu\tau}$. From the left panel of the figure, it is
evident that $\eps_{\tau\tau}$ has to be of $\mathcal O(10)$ to
significantly alter the prediction for $\eps_{\mu\tau}$. In the rest
of this work (including \fig~\ref{fig:combination}), the prior
put on $\eps_{\tau\tau}$ is $|\eps_{\tau\tau}| < 1.9$ ($1\sigma$ confidence level).

\subsection{Sensitivity to $\boldsymbol{\eps_{\mu\tau}}$}

In \fig~\ref{fig:emt-sensitivity}, we show the predicted sensitivity of OPERA to the NSI parameter $\eps_{\mu\tau}$.
\capenv{figure}{fig:emt-sensitivity}{
	\begin{center}
		\includegraphics[width=0.75\textwidth,clip=true]{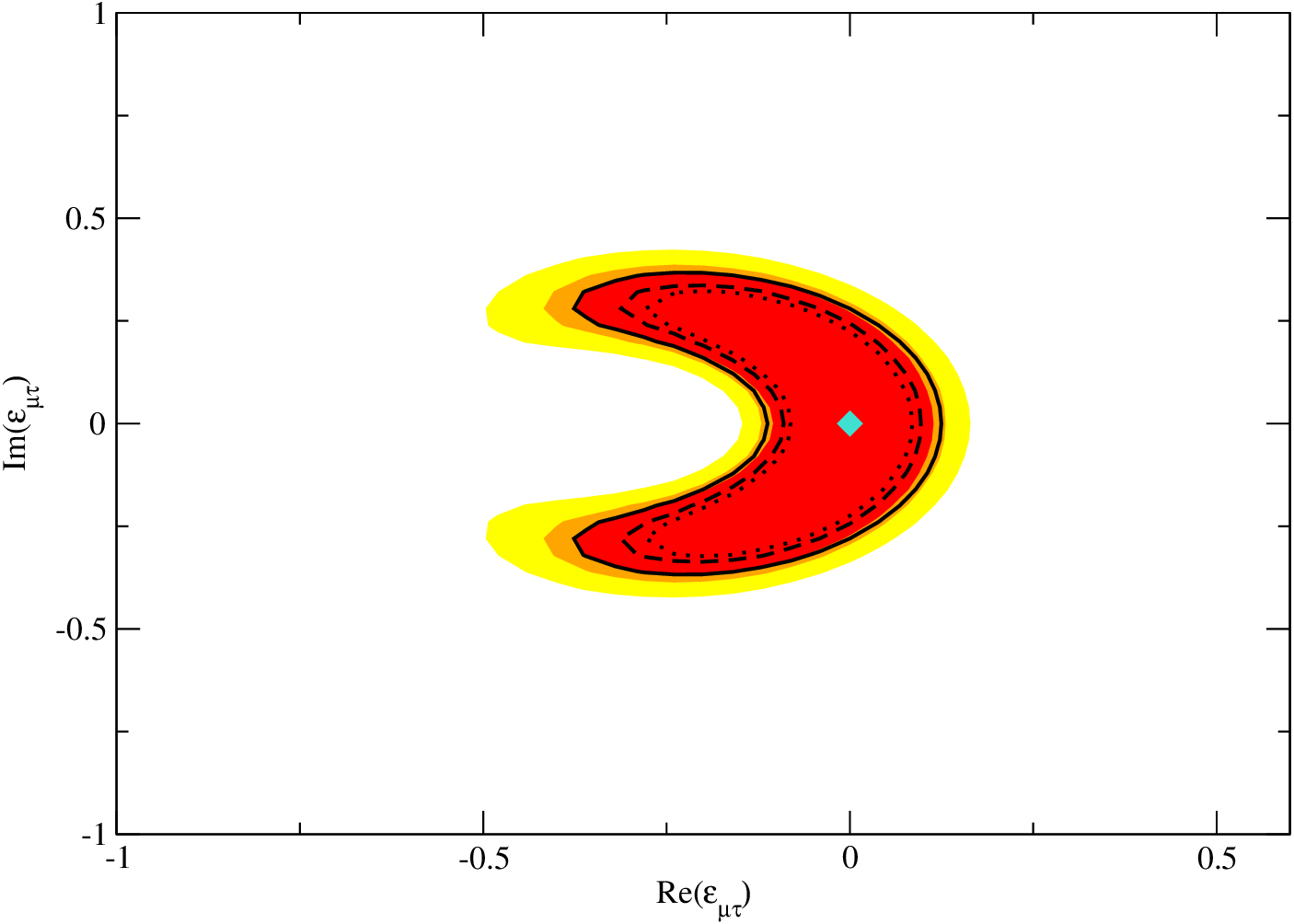}
	\end{center}%
}{%
The sensitivity of OPERA to the NSI parameter $\eps_{\mu\tau}$ (2~d.o.f.) for five (colored regions) and ten (curves) years of running
time. The simulated
value of $\eps_{\mu\tau}$ is zero and the confidence levels are 90~\%,
95~\%, and 99~\%, respectively. The
diamond corresponds to the simulated value of $\eps_{\mu\tau} = 0$.}
It is clear from this figure that the sensitivity contours extend in
the direction where the number of $\nu_\tau$ events is constant
(basically a circle centered at $\eps_{\mu\tau} \simeq
-0.3$). Furthermore, the sensitivity change if running the experiment
for a longer time has been indicated. As can be seen, this would
slightly improve the projected sensitivity. However, running the
experiment with reversed polarity could significantly improve the
sensitivity, see \fig~\ref{fig:polarity}.
\capenv{figure}{fig:polarity}
{%
\begin{center}
	\includegraphics[width=0.8\textwidth]{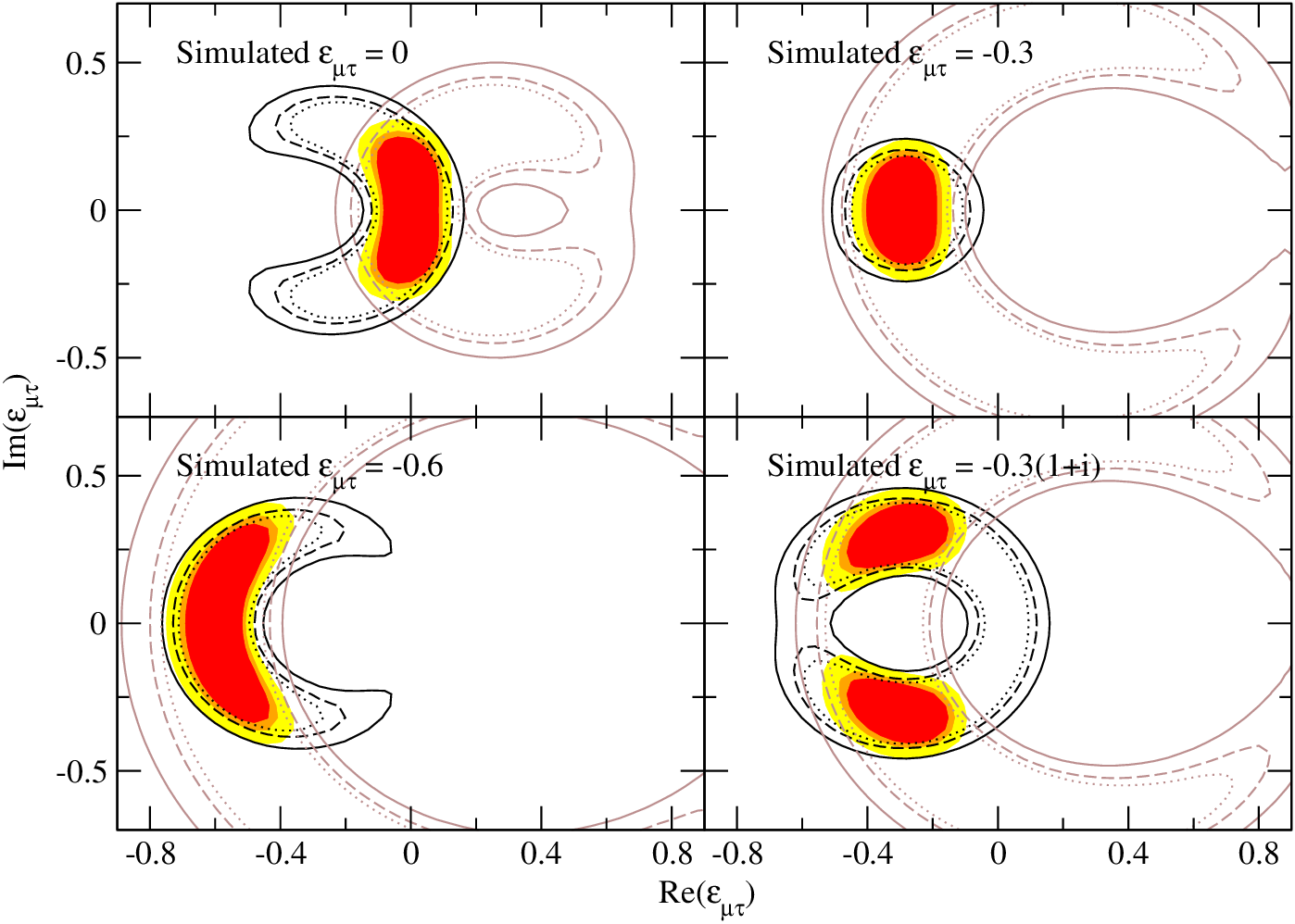}
\end{center}
}{%
The sensitivity of OPERA for five years of running time in neutrinos (dark
curves) and anti-neutrinos (light curves) as well as the combination
thereof (colored regions). The sensitivity levels correspond to
confidence levels (2~d.o.f.) of 90~\%, 95~\%, and 99~\%, respectively.}
Due to the change of $V \rightarrow -V$ and $\eps_{\alpha\beta}
\rightarrow \eps_{\alpha\beta}^*$ when considering anti-neutrinos
instead of neutrinos, it follows from \eq~(\ref{eq:pmutau}) that the
sensitivity contours for the reversed polarity will extend in a
different direction than those of the original polarity as can be
observed in the figure. Thus, the combination of the two polarities
could aid in resolving the degeneracy. That the anti-neutrino run by
itself produces slightly larger sensitivity contours is mainly due to the
lower cross-section. Furthermore, the figure shows the
effects of having different simulated values for $\eps_{\mu\tau}$.

In \fig~\ref{fig:non-zero:hierarchy}, the effects of fitting the data
to the wrong neutrino mass hierarchy are shown.
\capenv{figure}{fig:non-zero:hierarchy}{%
	\begin{center}
		\includegraphics[width=0.8\textwidth,clip=true]{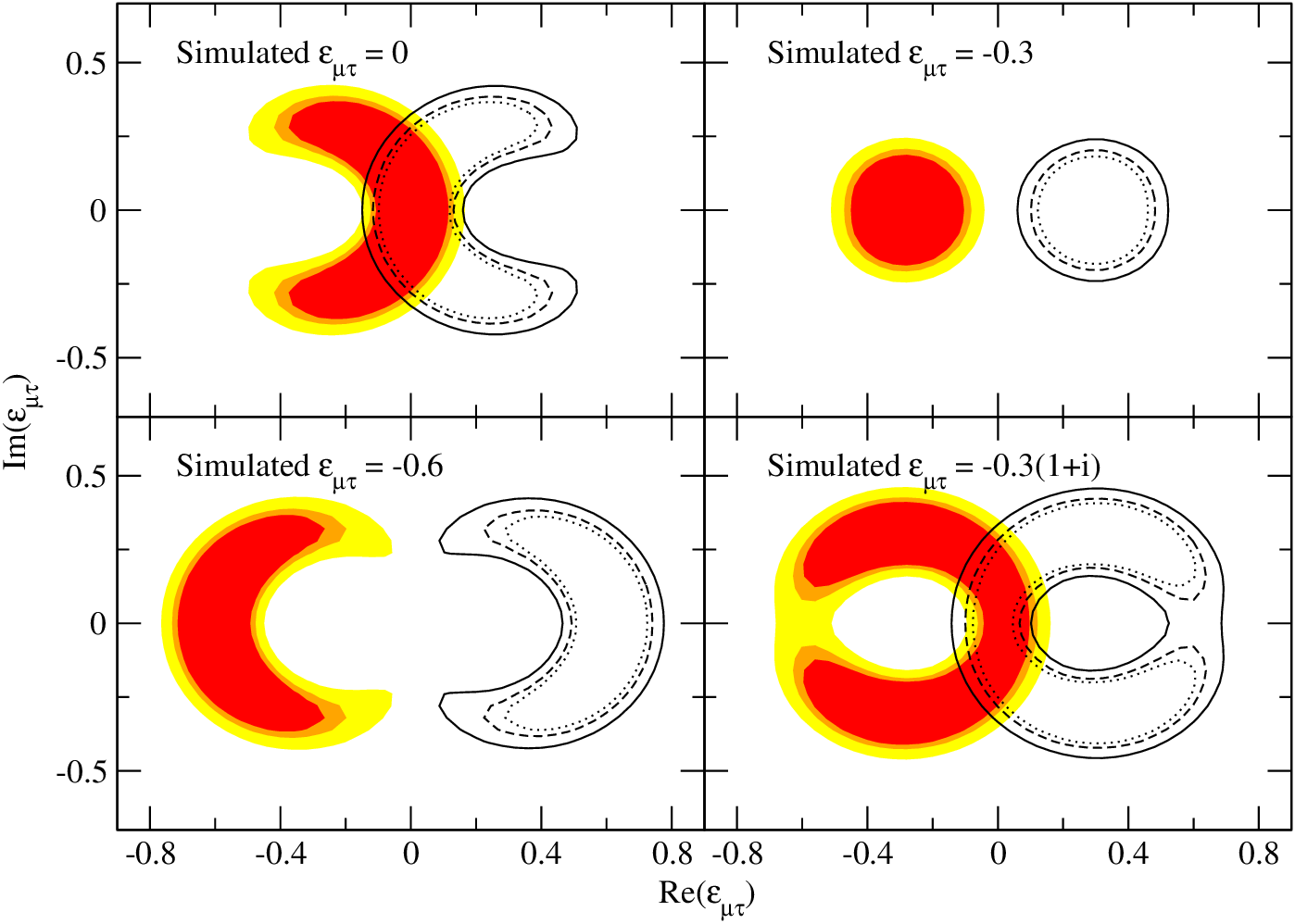}
	\end{center}%
}{%
The predicted sensitivity contours of OPERA for different
simulated values of $\eps_{\mu\tau}$ (2~d.o.f.). The simulated values
chosen are $\eps_{\mu\tau} = 0$ (upper-left panel), $\eps_{\mu\tau} =
-0.3$ (upper-right panel), $\eps_{\mu\tau} = -0.6$ (lower-left panel),
and $\eps_{\mu\tau} = -0.3(1+\I)$ (lower-right panel). The colored regions correspond to a
fit using the correct neutrino mass hierarchy, while the curves are the
regions obtained with a fit using the wrong neutrino mass
hierarchy. The confidence levels are 90~\%, 95~\%, and 99~\%,
respectively. This figure only includes the results of running for five years with neutrinos.}
Again, we can observe that the sensitivity contours extend in the
direction of a constant number of events, \ie, the circle centered at
$\eps_{\mu\tau} \simeq -0.3$. In the case of $\eps_{\mu\tau} = -0.3$,
the circle has radius zero and the allowed region is relatively
small. This corresponds to the anti-resonance case, where no events
are expected. The results of the fit using the wrong neutrino mass
hierarchy is a simple mirroring of the result with the correct
hierarchy, $\eps_{\mu\tau} \rightarrow -\eps_{\mu\tau}$. This can be
easily understood from \eq~(\ref{eq:pmutau}), where a sign change in
$\eps_{\mu\tau}$ exactly cancels the sign change in $\Delta m_{31}^2$
associated with changing the neutrino mass hierarchy. Strictly
speaking, the sensitivity of OPERA to $\eps_{\mu\tau}$ is the union of
the sensitivities obtained when fitting each mass hierarchy
separately.

\section{Summary and conclusions}
\label{concl}

We have studied NSIs in connection with the OPERA experiment. Unlike
in the previous work by Esteban-Pretel
\etal~\cite{EstebanPretel:2008qi}, where the focus was on the
effective NSI parameters $\eps_{e\tau}$ and $\eps_{\tau\tau}$ due to
the external bounds on the other parameters, we have focused on the
NSI parameter $\eps_{\mu\tau}$. The reason for this is that
$\eps_{\mu\tau}$ is more important for the $\nu_\mu \to \nu_\tau$
oscillation probability in OPERA due to the relatively short baseline,
as can be seen in our analytic considerations.

We have found that OPERA is actually sensitive to a combination of standard
and non-standard parameters, which can be easily observed in
\eq~(\ref{eq:pmutau}). The degeneracy in the parameter space, where
this combination is constant, is somewhat broken by the energy
dependence of the standard term. A much better determination of
$\eps_{\mu\tau}$ can be obtained if we consider a 5+5 year
neutrino-antineutrino run, especially if compared with a 10 year of
data taking with neutrinos only, the main reason being the different
correlations between standard and non-standard parameters, visible from
\eq~(\ref{eq:pmutau}) with the replacement $V \to -V$ and
$\eps_{\mu\tau} \to \eps^*_{\mu\tau}$.

We have also observed that the uncertainty on $\eps_{\mu\tau}$ can be
strongly worsened due to our ignorance in the sign of the large mass
squared difference $\Delta m_{31}^2$, resulting in a reflection of the allowed region for $\eps_{\mu\tau}$ in the imaginary axis. This holds true in the case of running in neutrinos only as well as running with both neutrinos and anti-neutrinos.

By means of the simple result of \eq~(\ref{eq:pmutau}), we found that
an \emph{anti-resonance} occurs when the standard and NSI parameters
cancel. In this case, the effective Hamiltonian element $H_{\tau\mu}$
vanishes and no $\nu_\mu \to \nu_\tau$ events would be observed.

The above analytic considerations were illustrated by our numeric
simulations using a modified version of the GLoBES software. In
particular, it is evident from \fig~\ref{fig:polarity} that running
the OPERA experiment for five years in each polarity would be much
more efficient in constraining the $\eps_{\mu\tau}$ parameter space
than running for ten years with neutrinos only.

Finally, we again want to mention that atmospheric neutrino experiments put constraints on
$\eps_{\mu\tau}$ which are better than what OPERA is sensitive to
\cite{Fornengo:2001pm}. However, these constraints have been computed
in a pure two-flavor framework without the interference of
$\eps_{e\tau}$. It is known that the bounds on $\eps_{\tau\tau}$ from
similar considerations are significantly weakened when extending to a
full three-flavor framework. Thus, OPERA will provide a very clean and
complementary bound, since only $\eps_{\mu\tau}$ enters into the
leading term of the neutrino oscillation probability $P_{\mu\tau}$ due
to the short baseline.

\section*{Acknowledgments}

We would like to thank Mark Rolinec and Walter Winter for useful
information about the GLoBES software and for the original AEDL file for describing the OPERA
experiment.

This work was supported by the Swedish Research Council
(Vetenskapsr{\aa}det), contract nos.~621-2005-3588 [T.O., M.B.] and
623-2007-8066 [M.B.], the Royal Swedish Academy of Sciences (KVA)
[T.O.], and the G\"oran Gustafsson Foundation [D.M.]. M.B.~and
D.M.~are grateful to the Royal Institute of Technology (KTH) for kind
hospitality during the development of this work.

\end{document}